\documentclass[runningheads]{llncs}
\usepackage{graphicx}
\usepackage{subcaption}
\usepackage{xcolor}
\usepackage{cite}
\begin{document}
%

\title{COVID-19’s Unequal Toll: An assessment of small business impact disparities with respect to ethnorace in metropolitan areas in the US using mobility data}
%
%
%
\author{Saad Mohammad Abrar \and
Kazi Tasnim Zinat \and
Naman Awasthi \and
Vanessa Frias-Martinez 
}
%

%
\authorrunning{SM Abrar et al.}
\titlerunning{Small Business Disparities by Ethnorace using mobility data}
%
\institute{University of Maryland at College Park \\
\email{\{sabrar,kzintas,nawasthi,vfrias\}@umd.edu}}
\maketitle              
\begin{abstract}
Early in the pandemic, counties and states implemented a variety of non-pharmacological interventions (NPIs) focused on mobility, such as national lockdowns or work-from-home strategies, as it became clear that restricting movement was essential to containing the epidemic. Due to these restrictions, businesses were severely affected and in particular, small, urban restaurant businesses. In addition to that, COVID-19 has also amplified many of the socioeconomic disparities and systemic racial inequities that exist in our society. 
The overarching objective of this study was to examine the changes in small urban restaurant visitation patterns following the COVID-19 pandemic and associated mobility restrictions, as well as to uncover potential disparities across different racial/ethnic groups in order to understand inequities in the impact and recovery. Specifically, the two key objectives were: 1) to analyze the overall changes in restaurant visitation patterns in US metropolitan areas during the pandemic compared to a pre-pandemic baseline, and 2) to investigate differences in visitation pattern changes across Census Block Groups with majority Asian, Black, Hispanic, White, and American Indian populations, identifying any disproportionate effects. Using aggregated geolocated cell phone data from SafeGraph, we document the overall changes in small urban restaurant businesses' visitation patterns with respect to racial composition at a granularity of Census Block Groups. 
Our results show clear indications of reduced visitation patterns after the pandemic, with slow recoveries. Via visualizations and statistical analyses, we show that reductions in visitation patterns were the highest for small urban restaurant businesses in majority Asian neighborhoods.

\keywords{COVID-19 business impact \and Mobility data \and Ethnoracial analysis}
\end{abstract}
\section{Introduction}
The outbreak of the COVID-19 pandemic brought unprecedented challenges to communities worldwide, prompting the implementation of various non-pharmaco-
logical interventions (NPIs) aimed at limiting mobility and curbing the spread of the virus \cite{baker2020impact, kuo2020collateral}. These interventions, such as nationwide lockdowns and remote work strategies, underscored the crucial role of restricting movement in containing the epidemic \cite{oka2021effect, oh2021mobility}. While these measures were effective in mitigating the health crisis, they also had significant repercussions for businesses, particularly small establishments that form the backbone of local economies~\cite{elimam2017role,cowling2015really}. 

The pandemic has also exposed and magnified pre-existing socioeconomic disparities and systemic inequities within our society. In fact, vulnerable populations, including racial and ethnic minorities, have borne a disproportionate burden of the pandemic's impact including anti-Asian racism ~\cite{huang2023cost} as well the condition of being frontline workers~\cite{oecd2022unequal}.
Understanding the impact of COVID-19 on small businesses and its intersection with racial composition is also vital for policymakers, community leaders, and stakeholders. By identifying disparities and informing targeted interventions, tangible work can be done towards fostering a more equitable recovery and supporting the long-term sustainability of small businesses.

To better understand the implications of COVID-19 on small businesses and to shed light on the intersection of ethnorace and business resilience, this study leverages aggregated geolocated cell phone data from SafeGraph, characterizing visits to small urban businesses in the US, with demographic data related to the social composition of the areas where these businesses are located. Analyzing business visitation patterns at the level of Census Block Groups in the US, we aim to provide insights into the changes in small urban business activity and their association with the race and ethnicity of their tracts.  
Although many types of businesses were affected by the pandemic, we focus on non-chain urban restaurants because these small businesses are major contributors to job creation while also being 
more vulnerable to falling demand~\cite{cowling2015really}. 

The objective of this research is twofold.
Firstly, we examine the overall changes of restaurant visitation patterns following the onset of the pandemic, assessing the extent to which small urban restaurant businesses in the US
have been affected and are recovering from the initial shock. Secondly, we investigate whether disparities in visitation patterns exist across different racial and ethnic groups, seeking to uncover potential inequities in the recovery process across metropolitan areas. By delving into the visitation patterns and changes in visitation volumes across restaurants, we can gain valuable insights into the specific challenges faced by small urban restaurant businesses in the US during the COVID-19 pandemic. This information will contribute to a more nuanced understanding of the restaurant industry's dynamics and inform targeted strategies and support mechanisms for these vulnerable businesses.



The rest of the paper is organized as follows. In Section 2,  we present a comprehensive description of the mobility and census data we will use in the analysis. Section 3 describes the  statistical and visualization methods we use to explore changes in restaurant visitation patterns during the pandemic, and to examine the relationship between ethnorace and visitation changes as well as recovery patterns. Finally, Section 4 will present our main findings, followed by Section 5 with a description of the proposed future work. 

\section{Data}
\subsection{Mobility Data}
The mobility patterns of millions of anonymized users in the US were tracked using mobile app software development kits (SDKs) by SafeGraph, a data provider that made this information publicly available at the beginning of the pandemic. To analyze the impact of COVID-19 on business visitation patterns, we utilized the points of interest (POIs) dataset from SafeGraph. We focused on two components of the POI data: 
\begin{itemize}
    \item \textbf{Core POI: } It includes the name and brand (indicating whether the POI is a part of a chain)
    \item \textbf{Patterns: } It includes the volume of daily visits per identified POI in the Safegraph database.
\end{itemize} 

To focus specifically on small business restaurants, our analysis involves extracting points of interest (POIs) that meet two criteria: they are not associated with any chain (brand), and they fall within the category of \textit{``Restaurants and Other Places"}. 
Around 80\% of POIs in SafeGraph have no brand associated as they are single commercial locations.
The visit patterns were measured as changes in volumes compared to the traffic before the pandemic. 
We collected SafeGraph data from January 1st, 2020 to January 31st, 2021.
The data collected between January 1st, 2020, and March 7, 2020, was defined as the baseline to assess changes in visitation patterns during the pandemic, which we mark with a starting date of March 15, 2020.
To reveal long-term trends that may be concealed by occasional fluctuations we utilized a 7-day rolling average to smooth the time series data. This approach enables us to identify significant changes and discern overall changes in the data, providing a clearer picture of the long-term effects of the pandemic on small business restaurant visitation patterns.

\subsection{ACS Data}
To evaluate disparities in the impact of COVID-19 on small urban restaurant businesses with respect to race and ethnicity, we utilized three datasets obtained from the American Census Survey datahub\footnote{\url{https://data.census.gov/}}. These datasets provided valuable information for our research:

\begin{itemize}
    \item \textbf{CBG ShapeFiles:} We utilized CBG ShapeFiles to gather the boundaries of the Census Block Groups. We use CBG as the geographical unit of analysis, assigning to each CBG the changes in volumes of visits to small restaurants located within the CBG. 
    
    \item \textbf{Urban-rural classification:} This dataset helped us identify urban CBGs by categorizing regions based on their urban level. We specifically focused on CBGs classified as level 1, which corresponds to large metropolitan counties. 
    
    \item \textbf{Race Distribution:} This dataset provided detailed information on the racial composition of the population at the Census Block Group (CBG) level. It allowed us to examine the distribution of different racial groups within specific geographic areas. To carry out the proposed analysis, each CGB was associated to the majority group race i.e., the race with the largest population volume in that CBG.

\end{itemize}

\begin{figure}
    \centering
    \includegraphics[scale=0.25]{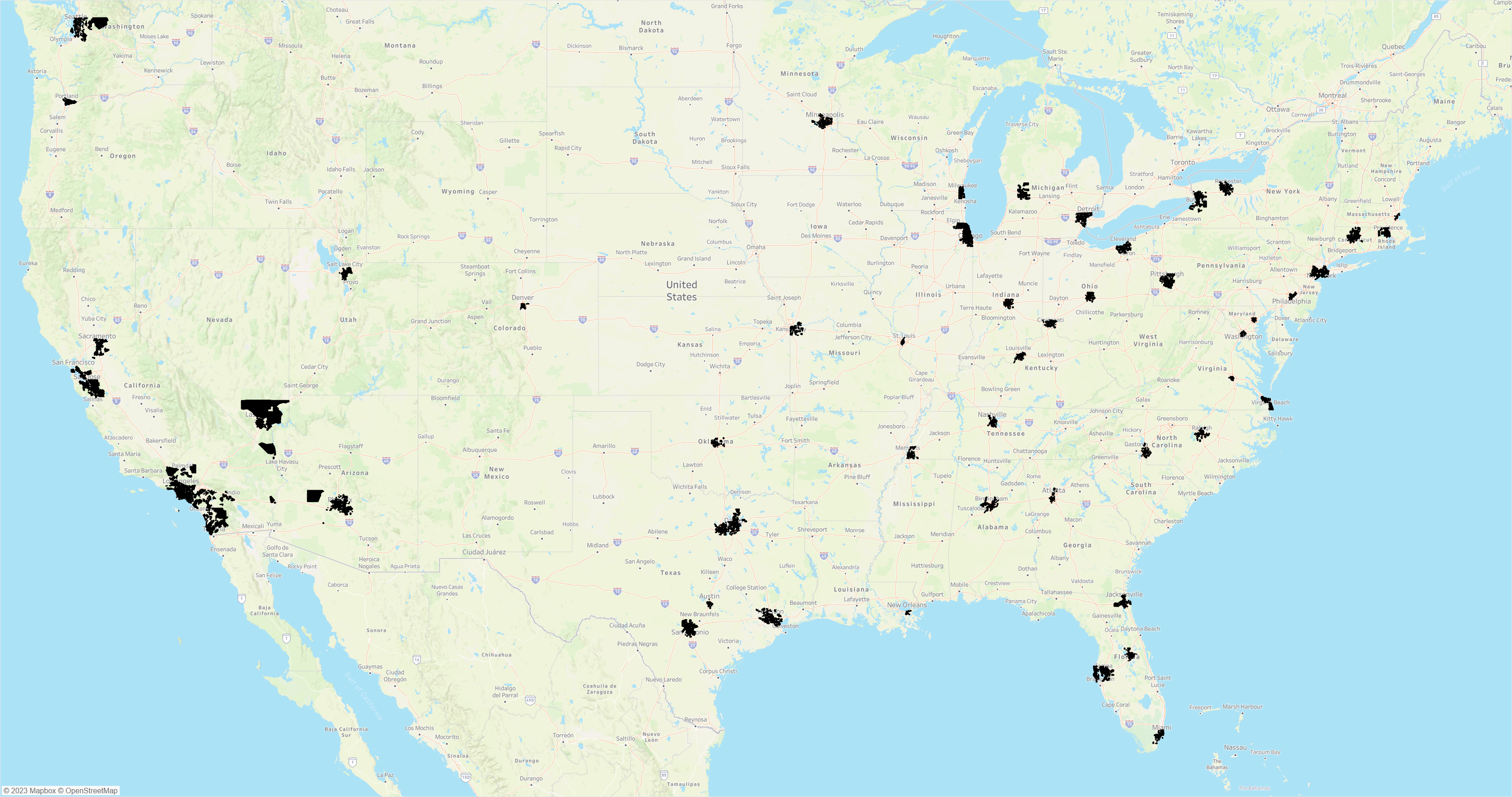}
    \caption{Urban Census Block Groups representing all the large metropolitan areas used in our analysis.}
    \label{fig:hotspots}
\end{figure}

\begin{figure}[!htp]
    \centering
    \includegraphics[scale=0.5]{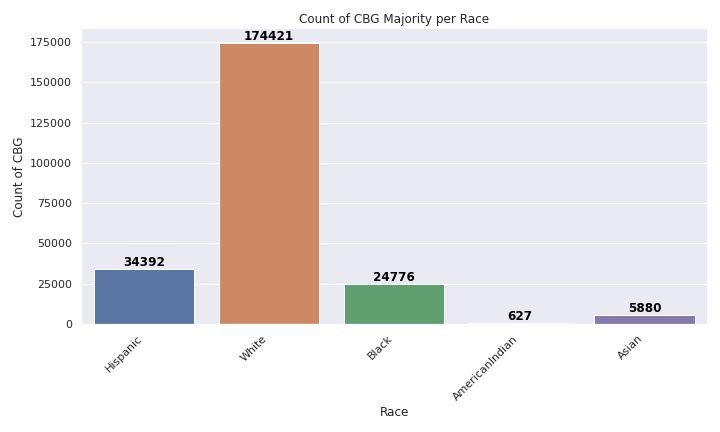}
    \caption{Count of CBGs associated to each majority race.}
    \label{pctCount}
\end{figure}

\begin{figure}[htp!]
\centering
\begin{subfigure}{0.49\textwidth}
    \centering
    
    \includegraphics[width=\linewidth]{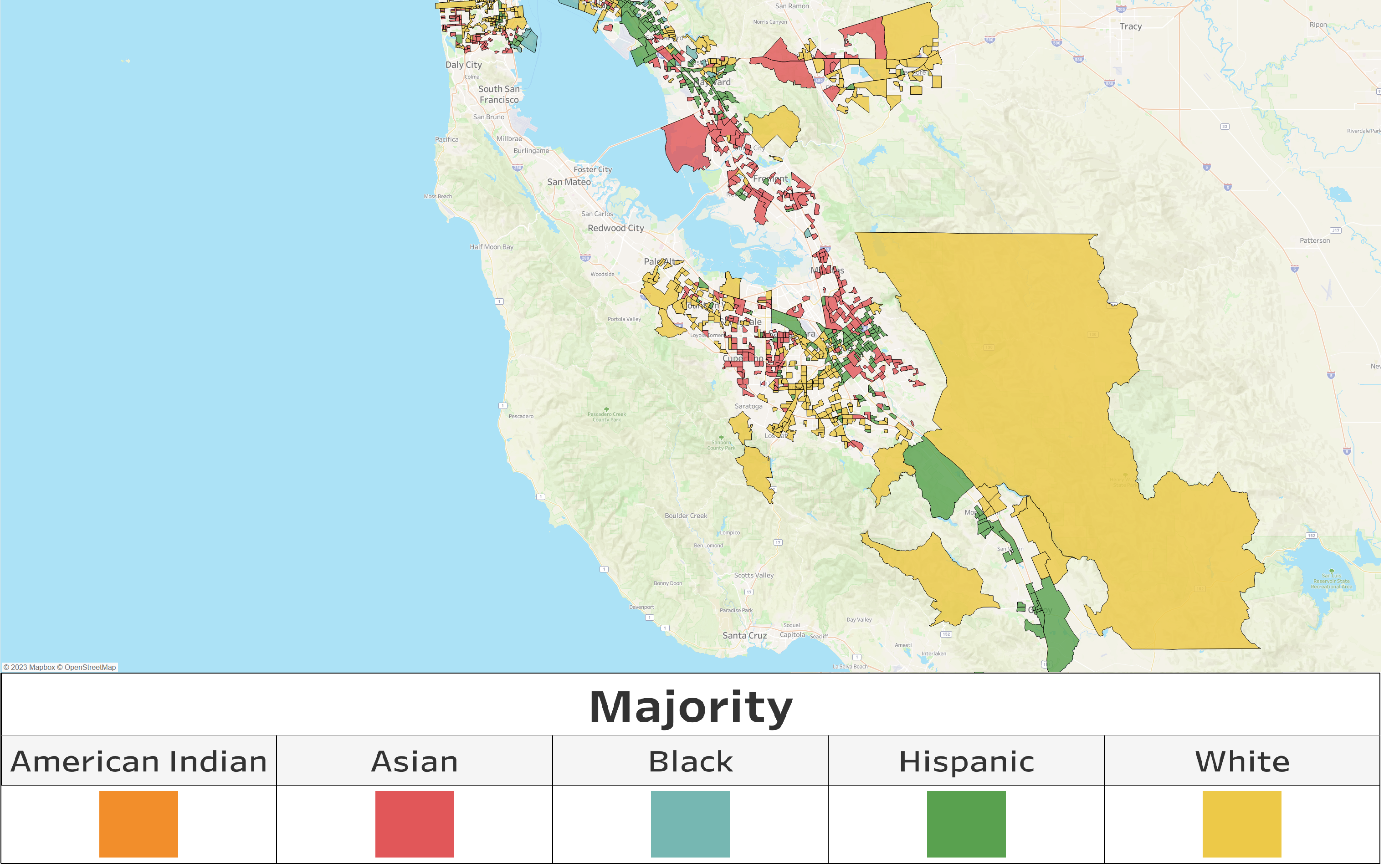}
    \caption{Racial composition of CBGs in the Bay Area, Northern California.} 
    \label{fig:ba}
\end{subfigure}
\hfill
\begin{subfigure}{0.49\textwidth}
    \centering
    \includegraphics[width=\linewidth]{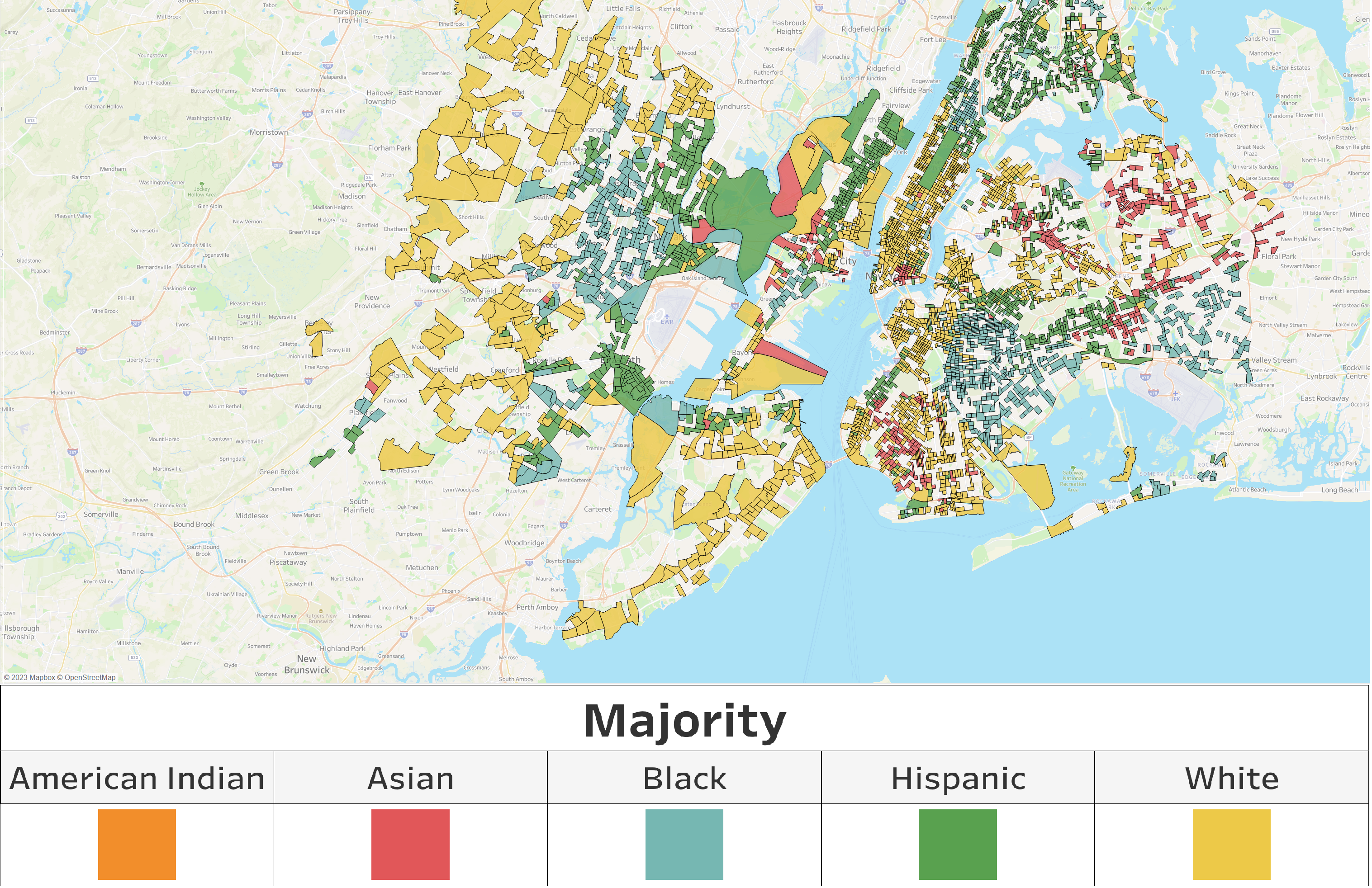}
    \caption{Racial composition of CBGs in and around New York City.}
    \label{fig:nyc}
\end{subfigure}

\caption{Racial composition at selected urban locations.}
\label{fig:racespatial}
\end{figure}

Figure \ref{fig:hotspots} shows all the urban Census Block Groups representing large metropolitan areas, providing an overview of the locations used in our analysis. Figure \ref{pctCount} represents the count of CBGs associated to each majority race. To delve deeper into the analysis, Figure \ref{fig:racespatial} shows an example of the racial composition across CBGs for the Bay area and for New York City.

\subsection{Why CBGs as a proxy measure of localized impact?}
Firstly, CBG allows for a localized impact assessment. It provides a finer level of granularity than larger geographic units, such as census tracts or counties. By focusing on CBGs, researchers can zoom in on specific neighborhoods or areas within a city or region. This localized approach enables a more precise understanding of the effects of interventions, policies, or demographic changes on smaller-scale populations.

Secondly, CBG is particularly useful when studying highly homogenous populations. In some cases, neighborhoods or communities may exhibit a high degree of homogeneity in terms of race, ethnicity, income level, or other demographic characteristics. By analyzing CBGs within these areas, researchers can gain insights into the dynamics and changes occurring within these homogeneous populations. This level of analysis can help identify trends, patterns, or disparities that may not be evident when examining larger, more heterogeneous geographic units.

Lastly, the reliance on CBGs as the unit of analysis is prevalent in most quantitative studies of neighborhood racial change \cite{brough2021understanding, ong2020covid, reibel2011neighborhood}. Researchers often utilize CBGs because they offer a balance between geographic precision and sample size. CBGs generally have a population size that is manageable for analysis while still providing enough data points to draw statistically significant conclusions. This widespread usage of CBGs as a unit of analysis allows for comparability and consistency across studies, facilitating the accumulation of knowledge and the identification of broader trends.


\section{Methodological Approach}

For each urban CBG in the US, we compute a time series with the daily percentage change in aggregated visits to small restaurants within that CBG. The daily percentage change reflects changes in volumes of visits during the pandemic with respect to the predefined pre-pandemic visitation patterns.
To investigate the recovery patterns across ethnoracial groups, we carry out separate analyses for each race and ethnicity by exclusively looking into the daily percentage change time series of census block groups with the same ethnoracial majority. For instance, to examine the pattern specific to the Asian demographic subgroup, we focus solely on the census clock groups - and their corresponding daily visitation time series - where Asians constitute the majority population.

The proposed methodological approach will allow us to examine overall changes in restaurant visitation patterns following the pandemic, and
whether disparities in visitation patterns exist across racial and ethnic groups, seeking to uncover potential inequities in the recovery process across metropolitan areas.
Nevertheless, a potential limitation of this approach is that it will not reveal the reasons behind changes in the visitation patterns \textit{i.e.,} small urban restaurants in specific ethnoracial CBGs could have received fewer visits during the pandemic either because residents of that CBG visited less their local restaurants, or because non-residents were visiting less that CBG. While the former would point to the impact of a specific ethnorace's visiting behaviors to their local restaurants; the latter would signal changes in visiting behaviors from non-residents to specific ethnoracial CBGs. 
In addition, changes in visitation patterns for specific ethnoraces could also be due to higher volumes of business closings. 

To evaluate overall small urban restaurants' recovery patterns, and changes across ethnoracial groups, we will also analyze the average percentage change in volumes of visits during the pandemic with respect to their corresponding pre-pandemic baselines, and we carry out this analysis for each ethnoracial group across three phases: phase I or social distancing (from mid March till mid May), phase II or management of the pandemic post-onset (from mid May till the end of November), and phase III or vaccination phase, to characterize the period when vaccines became available for certain social groups (from November onwards).

To evaluate statistical differences in recovery patterns across ethnoracial groups, we compute a 
two-sample Kolmogorov-Smirnov test~\cite{marozzi2013nonparametric} between each pair of ethnoracial average percentage change time series, with each average time series computed as the average of all time series across CBGs of the same majority ethnorace. 
 The KS test, which is non-parametric, is used to capture statistically significant similarities and differences in both location and scale between pairs of ethnoracial time series.
This focused approach allows us to analyze and compare the recovery trajectories among various racial communities. By employing this method, our objective is to identify any disparities or variations in 
post-pandemic recovery patterns based on racial composition within urban settings.



\section{Results}
\begin{figure}[htp!]
\centering
\begin{subfigure}{0.49\textwidth}
    \centering
    \includegraphics[width=\linewidth]{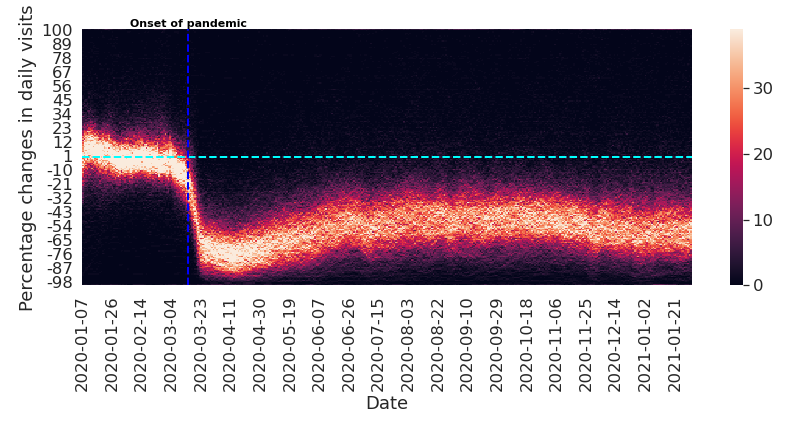}
    \caption{Heatmap matrix for CBGs across all groups} 
    \label{fig:ba}
\end{subfigure}
\hfill
\begin{subfigure}{0.49\textwidth}
    \centering
    \includegraphics[width=\linewidth]{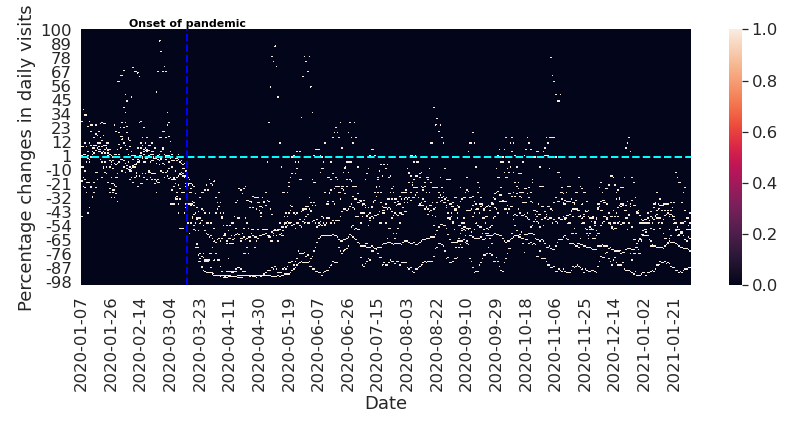}
    \caption{Heatmap matrix for CBGs with American Indian Majority}
    \label{fig:ai}
\end{subfigure}
\hfill
\begin{subfigure}{0.49\textwidth}
    \centering
    \includegraphics[width=\linewidth]{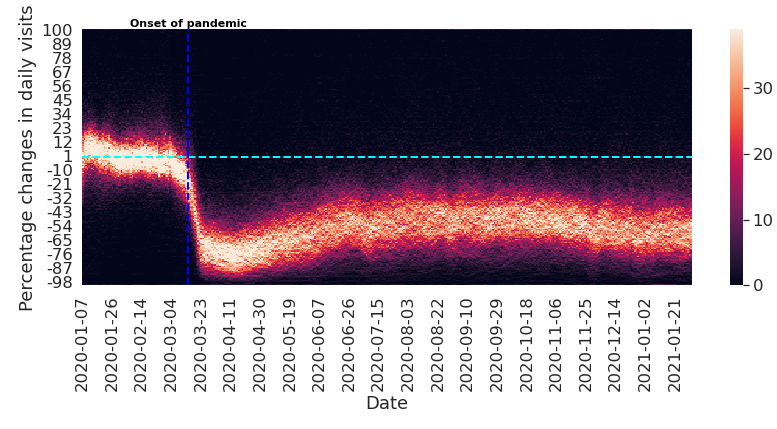}
    \caption{Heatmap matrix for CBGs with Asian Majority}
    \label{fig:asia}
\end{subfigure}
\hfill
\begin{subfigure}{0.49\textwidth}
    \centering
    \includegraphics[width=\linewidth]{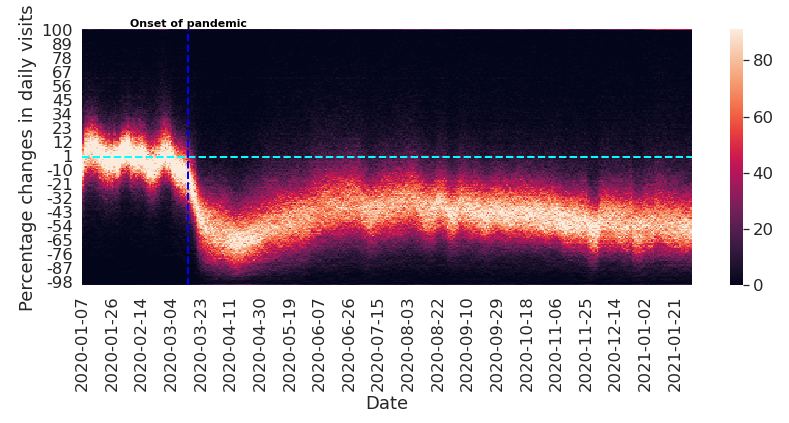}
    \caption{Heatmap matrix for CBGs \\ with Black Majority} 
    \label{fig:blck}
\end{subfigure}
\hfill
\begin{subfigure}{0.49\textwidth}
    \centering
    \includegraphics[width=\linewidth]{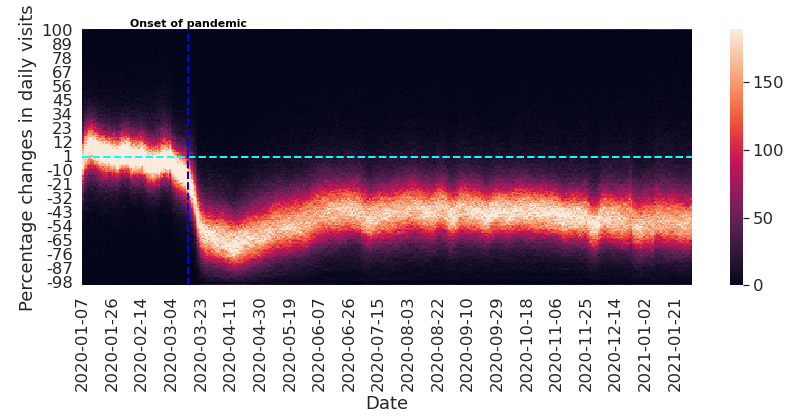}
    \caption{Heatmap matrix for CBGs with Hispanic Majority}
    \label{fig:his}
\end{subfigure}
\hfill
\begin{subfigure}{0.49\textwidth}
    \centering
    \includegraphics[width=\linewidth]{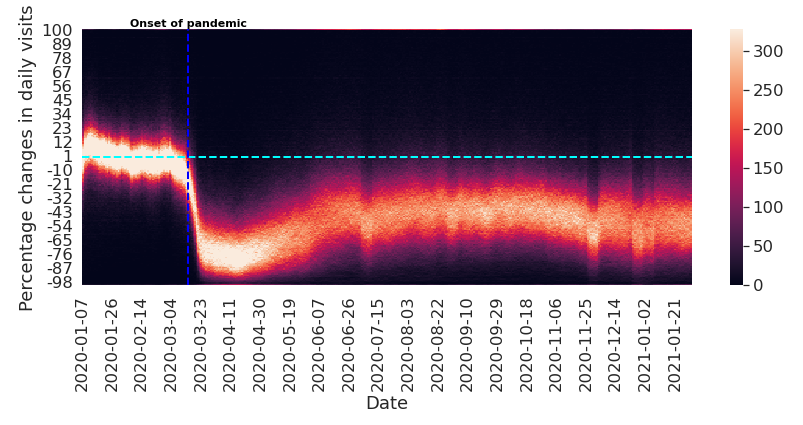}
    \caption{Heatmap matrix for CBGs with White Majority} 
    \label{fig:ba}
\end{subfigure}
\hfill

\caption{Heatmap matrices for percentage changes in visit patterns}
\label{fig:heatmap}
\end{figure}

\begin{figure}
    \centering
    \includegraphics[width = \linewidth]{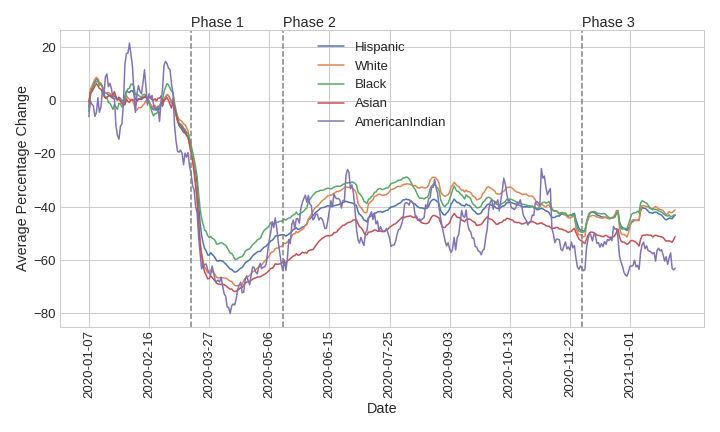}
    \caption{Average percentage change time series across all CBGs for each ethnicity or race. The vertical dashed lines separate the pre-pandemic phase from phases 1, 2 and 3.}
    \label{fig:race_avg}
\end{figure}

Figure \ref{fig:heatmap} shows heatmap matrices computed with the percentage change time series for the CBGs in our analysis. 
Each CBG is represented by a time series from January 7, 2020 to January 31, 2021; and every CBG contributes to a single pixel per day in the heatmap matrix.
The heatmap matrix's density reflects the combined contribution of each CBG, enabling a comprehensive visual representation of the variations in visitation patterns over time and across ethnoraces.
While \ref{fig:heatmap}(a) displays the heatmap matrix for all the CBGs, Figures \ref{fig:heatmap}(b--f) show
the percentage change time series represented as heatmap matrices for each ethnicity or race in our analysis.  
Figure ~\ref{fig:race_avg}, on the other hand, represents the average percentage change time series across all CBGs with the same majority race. In other words, for each heatmap matrix in Figures \ref{fig:heatmap}(b--f), we compute its average time series, which allows for a straight forward comparison of changes in restaurant visitation patterns across ethnoraces. 


As we can observe in Figure \ref{fig:heatmap}, right after the global onset of the pandemic (March 15, 2020) the overall volume of visits to small urban restaurants as well as the volumes of visits disaggregated by ethnorace heavily decreased and did not go back to normal even after one year into the pandemic. 
To analyze disparities in visitation patterns and trajectory recoveries across ethnoraces, Figure ~\ref{fig:race_avg} shows one time series per ethnorace, representing the average percentage changes in visitation patterns for small restaurants located in CBGs of a specific ethnorace. Overall, we observe that all CBGs, independently of their majority race were heavily affected by the pandemic with large percentage decreases of visitation patterns when compared to the baseline. We also observe that all time series reflect smooth trends across phases except for the American Indian time series, that suffers from more oscillations probably due to the smaller number of majority CBGs of that race in our dataset. Despite the large pandemic impact on visitation patterns, there exist nuanced differences across ethnoraces, with majority Asian and majority American Indian being not only the most impacted, but also the ones with the slowest recovery even after one year into the pandemic.

To quantify these differences, Table ~\ref{visitsByPhase} displays the average percentage changes in visitation patterns per ethnorace and across the three phases explained earlier. We can observe that Asian and American Indian CBGs are the ones most impacted, with larger average percentage decreases in visits with respect to pre-pandemic times 
of $42.7\%$ and $41.7\%$, respectively, across all phases; reaching maximum decreases of $63\%$ and $62\%$ respectively for phase 1 (onset of the pandemic). Table ~\ref{kstest} shows the KS statistics and p-values for all combinations of ethnoracial average percentage change time series. 
The null hypothesis was rejected for all pairs of distinct time series ($p\_value = 0$) and the Asian average percentage change time series is the one most different to the others (with the highest ks statistic values). Put together with the findings in Table ~\ref{visitsByPhase}, this statistical analysis shows that Asian CBGs are not only the least visited across pandemic phases, but that they are also the most statistically significantly different from the other ethnorace's distributions, with ks statistics in the $0.48-0.58$ range when compared to majority Hispanic, White and Black CBGs, albeit more similar to the American Indian time series ($ks=0.22$).

\begin{table}[!htp]\centering

\resizebox{\textwidth}{!}{\begin{tabular}{c|c|c|c|c|c}
 & American Indian & Asian & Black & Hispanic & White \\
\hline
\hline
Phase 1 &-61.9199 (11.42972) &-63.0853 (11.52541) &-49.3982 (9.574766) &-54.6981 (10.45263) &-60.1125 (11.40951)\\
Phase 2 &-44.7563 (8.186457) &-47.9976 (4.229237) &-36.8484 (4.287543)  &-41.2918 (3.107061) &-37.1945 (5.706389)\\
Phase 3 &-57.1392 (4.839321) &-51.597 (1.347966) &-43.0362 (2.738588) &-44.043 (2.244181) &-43.8665 (3.330658) \\
All Phases &-41.7721 (21.8725) &-42.709 (20.79114) &-33.4509 (16.75843) &-36.7284 (18.04656) &-35.4661 (19.03601) \\
\hline
\end{tabular}}
\caption{Average percentage change and the standard deviation (std) in visitation patterns per ethnorace over all the time period under study as well as for each phase.}
\label{visitsByPhase}
\end{table}

\begin{table}[!htp]
\centering
\resizebox{\textwidth}{!}{\begin{tabular}{c|c|c|c|c|c}
\textbf{KS\_Test} & American Indian &Asian &Black &Hispanic &White \\
\hline
\hline

American Indian &(0.0, 1.0) &(0.21995, 0.0) &(0.38107, 0.0) &(0.30179, 0.0) &(0.31969, 0.0) \\
Asian &(0.21995, 0.0) &(0.0, 1.0) &(0.58312, 0.0) &(0.48593, 0.0) &(0.50639, 0.0) \\
Black &(0.38107, 0.0) &(0.58312, 0.0) &(0.0, 1.0) &(0.27877, 0.0) &(0.10997, 0.01761) \\
Hispanic &(0.30179, 0.0) &(0.48593, 0.0) &(0.27877, 0.0) &(0.0, 1.0) &(0.32481, 0.0) \\
White &(0.31969, 0.0) &(0.50639, 0.0) &(0.10997, 0.01761) &(0.32481, 0.0) &(0.0, 1.0) \\
\hline
\end{tabular}}
\caption{Two-sample Kolmogorov-Smirnov tests between each pair of ethnoracial average percentage change time series. For each pair of distributions we report the ks statistic and its significance (p-value).}
\label{kstest}
\end{table}

\section{Future Directions}
The initial results point to interesting insights with respect to impact and recovery visitation patterns across ethnoraces. Nevertheless, further analysis is necessary to obtain more conclusive results. 
Several approaches can be undertaken to enhance the analysis and derive meaningful insights. These include:
\begin{itemize}
    \item \textbf{Disaggregating CBGs by County:} Instead of analyzing all the CBGs together, and aggregate by ethnorace, conducting an analysis that focuses on individual counties can provide more localized and targeted insights. This approach allows for a deeper understanding of the variations and patterns within specific geographic regions. In addition to that, examining and reporting statistically disproportionate behaviors per county can shed light on any significant deviations from the expected patterns. Furthermore,  analysis can help identify counties where the impact on small urban restaurant businesses' visitation patterns significantly differ from others, indicating potential underlying factors influencing recovery.


    \item \textbf{Analysis of Policy Effects:} Analyzing the impact of different policies enacted during the pandemic on small urban restaurant visitation patterns can provide insights into how specific regulations influenced small businesses. This analysis can involve assessing the effects of measures like lockdowns, capacity restrictions, or curfews on the visitation trends and recovery trajectories.

    \item \textbf{Regression Analysis:} Conducting regression analysis on the percentage change in visits, with specific demographic variables as independent features, can help determine the relationships and influences between demographic factors and visitation patterns. Therefore, this analysis can provide a more nuanced understanding of how factors like racial composition, income levels, or education levels might have an effect on small urban business visitation patterns during and after the pandemic.

    \item \textbf{Causal Analysis of Changes in Visitation Patterns:} Besides the analysis of the statistical changes in visitation patterns presented, it would also be important to understand the causes behind the changes. Visitation patterns can change because internal CBG behaviors change, pointing to specific ethnoracial communities responding differently to the pandemic; because external CBG behaviors change, pointing to how other ethnoraces might change their visits to small restaurants located in specific ethnoracial CBGs; or because small restaurants located in CBGs of a specific ethnorace might be closing at higher rates. Identifying the role of all of these factors is critical to inform targeted strategies and support mechanisms for vulnerable businesses.

By incorporating these approaches into the analysis, we posit we can
obtain more conclusive and robust findings. These methods allow for a comprehensive examination of the data and will enhance our understanding of the factors influencing small urban restaurants' visitation patterns during the pandemic.

\end{itemize}

\bibliographystyle{splncs04}
\bibliography{ref}
%
\end{document}